\documentclass[aps,pre,reprint,groupedaddress,showpacs]{revtex4-1}
\usepackage{amsmath,amssymb, epsfig}

\begin{document}


\title{Bound on Thermoelectric Power in a Magnetic Field 
within Linear Response}


\author{Kay Brandner and Udo Seifert}

\affiliation{II. Institut f\"ur Theoretische Physik, Universit\"at Stuttgart, 70550 Stuttgart, Germany}


\date{\today}

\begin{abstract}
For thermoelectric power generation in a multi-terminal geometry,
strong numerical evidence for a universal bound as a function of
the magnetic-field induced asymmetry of the non-diagonal Onsager
coefficients is presented. This bound implies, inter alia, that the power
vanishes at least linearly when the maximal efficiency is approached.
In particular, this result rules out that Carnot efficiency can be
reached at finite power, which an analysis based on the second law
only, would, in principle, allow.
\end{abstract}


\maketitle

\newcommand{\vectt}[2]{\left(\!\begin{array}{c}#1\\#2\end{array} \!\right)}
\newcommand{\matt}[4]{\left(\!\begin{array}{cc} #1 & #2\\ #3 & #4 \end{array} \!\right)}
\newcommand{\F}{\mathcal{F}}
\newcommand{\T}{\mathcal{T}}
\newcommand{\e}{\varepsilon}
\newcommand{\abs}[1]{|#1|}

\section{Introduction}
Thermoelectric generators use a heat current between two reservoirs
of temperatures $T_{{\rm h}}$ and $T_{{\rm c}}<T_{{\rm h}}$ connected 
via a device to drive a particle current against a gradient in
chemical potential, thus delivering useful power.
As a fundamental consequence of the second law of thermodynamics, the 
efficiency $\eta$ of such a heat engine is bounded by the Carnot value 
$\eta_{{\rm C}}\equiv 1-T_{{\rm c}}/T_{{\rm h}}$.
This bound can indeed be reached if the device acts as a perfect 
energy filter \cite{Mahan1996}, however, only at the price of 
vanishing power \cite{Humphrey2002, Humphrey2005}.

The relationship between thermoelectric power and efficiency beyond
 this limiting case has been widely studied during the last 
years \cite{Esposito2009a, Leijnse2010, Nakpathomkun2010, Sanchez2013,
Kennes2013, Mazza2014}, see also \cite{Benenti2013} and 
\cite{Sothmann2014} for recent reviews. 
A particular landmark was achieved by Whitney 
\cite{Whitney2014, Whitney2014a}, who calculated the transmission 
function that leads to optimal efficiency at given power and thereby
identified upper bounds on the output power of quantum coherent devices. 
In all cases, however, finite power is only possible at an efficiency 
strictly smaller than $\eta_{{\rm C}}$.

This antivalence resembles the power-efficiency-dilemma, which has 
been studied by Allahverdyan et al. on the basis of a quite general
model for a re\-ciprocating quantum heat engine \cite{Allahverdyan2013}.
Their results confirm the general expectation that, under realistic 
conditions, a conventional cyclic heat engine can approach Carnot
efficiency only in the quasi-static limit, within which its power 
output necessarily goes to zero.

In strong contrast, Benenti \emph{et al.} have recently pointed 
out that a magnetic field, which breaks the symmetry between the 
off-diagonal Onsager coefficients, might enhance the efficiency of a 
thermoelectric heat engine substantially such that 
$\eta_{{\rm C}}$ seems to be achievable even at finite power 
\cite{Benenti2011}. 
This exciting suggestion follows from a straightforward analysis of 
the linear response regime using only the second law. 
However, so far, neither a specific model for a thermoelectric 
generator delivering finite power at Carnot efficiency, nor a
fundamental principle forbidding the existence of such a super device
has been discovered. 

A promising platform for the investigation of both of these aspects 
is provided by the multi-terminal setup sketched in Fig. 
\ref{FIG_Model}. 
By operating this model as a thermoelectric heat engine, we 
previously derived bounds on its efficiency $\eta$ that are stronger
than the ones obtained by Benenti \emph{et al.}
\cite{Brandner2013, Brandner2013a}.
These bounds, which follow from current conservation, depend only on
the number $n$ of terminals and, for any finite $n$, constrain $\eta$
to be strictly smaller than $\eta_{{\rm C}}$ whenever the off-diagonal 
Onsager coefficients are not identical thus preventing finite
power at Carnot efficiency.

In this paper, we move an essential step forward by investigating 
whether not only efficiency but also power can be bounded. 
Since power $P$ carries a physical dimension, such an endeavor can be 
expected to be harder, and, a priori, less universal, than searching
for a bound on the dimensionless efficiency. 
Bounding power is arguably more relevant since a high efficiency is 
useless from a practical point of view if it comes with minuscule 
power.
Here, we provide strong numerical evidence that $P$ is indeed subject 
to a new bound.
This bound implies that the power output of the device vanishes
whenever its efficiency approaches the upper limit corresponding to 
the respective number of terminals. 
Finally, by extrapolating our findings to arbitrary $n$, we derive a 
conjecture for a universal bound on $P$, which rules out the option of 
finite power at Carnot efficiency even in the 
limit $n\rightarrow\infty$.

The paper is organized as follows.
In section II, we give give a brief review of the essentials of the 
multi-terminal model and the bounds on its efficiency as a 
thermoelectric heat engine, which were derived in our previous work 
\cite{Brandner2013,Brandner2013a}.
Section III contains our main results. 
We outline the numerical procedure used to substantiate our new 
constraint on the Onsager coefficients and show how this constraint
can be used to bound the power of the multi-terminal thermoelectric 
generator. 
Finally, we conclude in section IV. 
\section{Multi-terminal model}

\subsection{Setup}
\begin{figure}
\epsfig{file=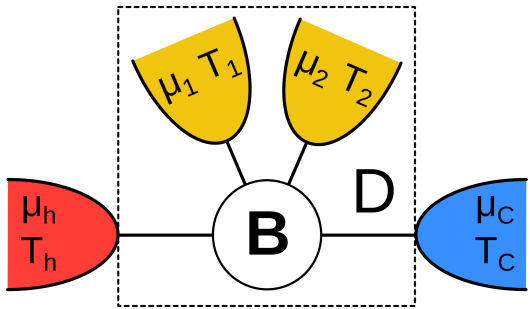,scale=0.13}
\caption{Sketch of a thermoelectric generator. 
Both, heat and particle current run from the hot reservoir on the 
left to the cold reservoir on the right thereby traversing the device
D, which is represented by the dashed box. 
Within the multi-terminal setup, the device consists of a central 
scattering region subject to a magnetic field $\mathbf{B}$ and 
$n-2$ probe terminals. 
The figure corresponds to $n=4$. 
\label{FIG_Model}}
\end{figure}
We consider the model sketched in Fig. \ref{FIG_Model}. 
On the primary level, $n$ electronic reservoirs of respective 
temperature $T_i$ and chemical potential $\mu_i$ $(i={{\rm h}},1,
\dots,n-2,{{\rm c}})$, are connected via perfect, one-dimensional 
leads to a central scattering region penetrated by the magnetic field 
$\mathbf{B}$. 
The particle and heat currents flowing from the reservoir $i$ to the 
scattering region are denoted by $J_\rho^i$ and $J_q^i$, the 
corresponding affinities by 
\begin{equation}
\F_\rho^i\equiv(\mu_i-\mu)/T
\quad\text{and}\quad 
\F_q^i\equiv(T_i-T)/T^2
\end{equation}
respectively, where the reference values 
$\mu=\mu_{{\rm c}}$ and $T=T_{{\rm c}}$ have been chosen. 
In the linear response regime, to which we will stick throughout the 
paper, the phenomenological equations describing the coupled heat and 
particle transport in this system read
\begin{equation}\label{Phenomenological_Equations}
\mathbf{J}^i = \sum_{j={{\rm h}},\dots,n-2}\mathbb{L}_{ij}
\boldsymbol{\F}^j
\end{equation}
with $\mathbf{J}^{{i}}\equiv (J_\rho^i,J_q^i)^t$, 
$\boldsymbol{\F}^i\equiv (\F^i_\rho,\F^i_q)^t$ and $\mathbb{L}_{ij} \in 
\mathbb{R}^{2\times 2}$ for $i,j = {{\rm h}}, 1,\dots,n-2$.
Since, at this stage, we assume uncorrelated particles, which 
move coherently through the leads and the scattering region, the 
matrix blocks $\mathbb{L}_{ij}$ can be calculated explicitly using the
multi-terminal Landauer formula \cite{Butcher1990, Sivan1986}
\begin{equation}\label{Landauer_Buettiker_Formula}
\mathbb{L}_{ij} \! =\!  \int_0^\infty \!\!\!\!\!\! dE\; f(E)
\left(\!\! \begin{array}{cc}
1 \!  & E-\mu\\ E-\mu \! & (E-\mu)^2
\end{array}\!\!\!\right)
\left( \delta_{ij}- \T_{ij}(E)\right) \!\!
\end{equation}
with $f(E)\equiv\cosh^{-2}\left((E-\mu)/(2k_B T)\right)/(4k_B h)$,
where $h$ denotes Planck's constant and $k_B$ Boltzmann's constant. 
The energy and magnetic field dependent transmission coefficients 
$\T_{ij}(E)$ encode the properties of the scattering region and cover
the effects of particle-particle interactions on the mean field level
\cite{Christen1996}.
Due to current conservation and time reversal symmetry, these 
dimensionless quantities must fulfill the sum rules 
\cite{Butcher1990, Sivan1986}
\begin{equation}\label{Sum_Rules_Transmission_Coefficients}
\sum_{i={{\rm h}},\dots,{{\rm c}}} \T_{ij}(E) 
= \sum_{j={{\rm h}},\dots,{{\rm c}}} \T_{ij}(E)= 1
\end{equation}
and the symmetry relation 
\begin{equation}\label{Symmetry_Transmission}
\T_{ij}(E,\mathbf{B}) = \T_{ji}(E,-\mathbf{B}),
\end{equation}
respectively.
Throughout the paper, we notationally suppress the dependence of any
quantity on $\mathbf{B}$, if there is no need to indicate it 
explicitly.
For vanishing magnetic field, relation (\ref{Symmetry_Transmission}) 
implies that the primary Onsager matrix showing up in 
(\ref{Phenomenological_Equations}) must be symmetric. 
For a fixed $\mathbf{B}\neq 0$, however, (\ref{Symmetry_Transmission})
does not lead to any further constraint. 

The full multi-terminal setup described above becomes a model for a 
thermoelectric heat engine by imposing the boundary conditions 
$J^i_\rho=J^i_q = 0$ for $i=1,\dots,n-2$.
Using these constraints to eliminate the corresponding affinities 
$\F^i_\rho,\F^i_q$, we obtain the reduced set of phenomenological 
equations 
\begin{equation}
\vectt{J_\rho^{{{\rm h}}}}{J_q^{{{\rm h}}}}
=\matt{L_{\rho\rho}}{L_{\rho q}}{L_{q\rho}}{L_{qq}}
\vectt{\F_\rho^{{{\rm h}}}}{\F_q^{{{\rm h}}}},
\end{equation}
where the $L_{\alpha\beta}$ with $\alpha,\beta = \rho, q$ denote 
effective Onsager coefficients, which are given by 
\begin{widetext}
\begin{equation}\label{Effective_Onsager_Matrix}
\left(\!\begin{array}{cc}
L_{\rho\rho} & L_{\rho q}\\
L_{q\rho} & L_{qq}
\end{array}\!\right)
= \mathbb{L}_{{{\rm hh}}}- \Bigl( \mathbb{L}_{{{\rm h}}1},\dots,
\mathbb{L}_{{{\rm h}}n-2} \Bigr)
\left(\!\! \begin{array}{ccc}
\mathbb{L}_{11} & \cdots & \mathbb{L}_{1n-2}\\
\vdots & \ddots & \vdots\\
\mathbb{L}_{n-21} & \cdots & \mathbb{L}_{n-2\;n-2}
\end{array}\!\right)^{\!\!-1}\!\!\!
\left(\!\begin{array}{c}
\mathbb{L}_{1{{\rm h}}}\\
\vdots\\
\mathbb{L}_{n-2{{\rm h}}}
\end{array}\!\right),
\end{equation}
\end{widetext}
with $\mathbb{L}_{ij}$ the blocks of primary Onsager coefficients  
defined in (\ref{Landauer_Buettiker_Formula}).

The physical intuition behind this procedure is that only the hot and
the cold terminal are considered as real, while the remaining ones act
as probes, which do not exchange any net quantities with the physical 
reservoirs.
Originally proposed by B\"uttiker \cite{Buttiker1986}, such probe 
terminals have meanwhile become a well established method to 
phenomenologically model inelastic scattering events like 
electron-phonon interactions, see, for example, \cite{Brandner2013,
Brandner2013a,Saito2011,Entin-Wohlman2012,Balachandran2013}.  
Such processes would be hard to take into account on a microscopic 
level, but are, inter alia, crucial to obtain a non-symmetric Onsager
matrix \cite{Buttiker1988a,Buttiker1986b}.

\subsection{Bounds on efficiency}

The thermodynamic efficiency of a thermoelectric generator in the linear
response regime generally reads 
\begin{equation}\label{Efficiency}
\eta \equiv \frac{P}{J_q^{{\rm h}}}= - \eta_{{{\rm C}}} 
\frac{L_{\rho\rho}\gamma^2+L_{\rho q}\gamma}{L_{q\rho}\gamma+L_{qq}},
\end{equation}
where 
\begin{equation}\label{Power_Output_General}
P=(\mu-\mu_{{\rm h}})J^{{\rm h}}_\rho
= -T(\F_q^{{{\rm h}}})^2 \left(L_{\rho\rho}\gamma^2 
+ L_{\rho q}\gamma\right),
\end{equation}
is the supplied power, 
\begin{equation}
\gamma\equiv\F_\rho^{{{\rm h}}}/\F_q^{{{\rm h}}}
\end{equation}
denotes the ratio of the affinities and $\eta_{{{\rm C}}}\equiv
(T_{{\rm h}}-T)/T=T\F_q^{{{\rm h}}}$ the linear response expression of
the Carnot efficiency. 
Maximizing $\eta$ with respect to $\gamma$ yields the maximum 
efficiency 
\begin{equation}\label{Max_Efficiency}
\eta_{{\rm max}}(x,y)\equiv
\eta_{{\rm C}} x \frac{\sqrt{y+1}-1}{\sqrt{y+1}+1}.
\end{equation}
Here, we introduced the dimensionless parameters 
\begin{equation}\label{xy_Definition}
x\equiv \frac{L_{\rho q}}{L_{q\rho}}
\quad\text{and}\quad
y\equiv 
\frac{L_{\rho q}L_{q\rho}}{L_{\rho\rho}L_{qq}-L_{\rho q}L_{q\rho}},
\end{equation}
which, due to the second law, have to obey the inequalities 
\cite{Benenti2011}
\begin{equation}\label{xy_Constraint_Second_Law}
h(x)\leq y\leq 0 \;\; \text{for} \;\; x<0,\;\;
0\leq y \leq h(x)\;\; \text{for} \;\; x>0
\end{equation}
with $h(x)\equiv 4x/(x-1)^2$.
For $\mathbf{B}=0$, $x$ assumes the symmetric value $1$ as 
a consequence of time reversal symmetry and $y$ reduces to the 
conventional figure of merit $ZT$ \cite{Benenti2011, Bell2008}.

In our previous work on the $n$-terminal setup \cite{Brandner2013a},
we showed that current conservation implies that 
(\ref{xy_Constraint_Second_Law}) can be strengthened by replacing 
$h(x)$ with
\begin{equation}
h_n(x)\equiv \frac{4x \cos^2(\pi/n)}{(x-1)^2}.
\end{equation}
The resulting constraint leads to the bound 
\begin{equation}\label{Max_Efficiency_Bound}
\eta_{{\rm max}}(x)\equiv \eta_{{{\rm C}}} x 
\frac{\sqrt{h_n(x)+1}-1}{\sqrt{h_n(x)+1}+1}
\end{equation}
on the efficiency as a function of the asymmetry parameter $x$, 
which, for any finite number of terminals, is stronger than the
one required by the bare second law. 
However, no predictions on a putative maximal power can be drawn from
the analysis presented so far. 

\section{Bounds on power}
The fundamental bound on power discussed in this paper follows 
from the constraint
\begin{equation}\label{Lyh_Bound}
\bar{L}_{qq}\equiv L_{qq}/N_{qq} \leq 1-y/h_n(x)
\end{equation}
on the Onsager coefficient $L_{qq}$, where
\begin{equation}\label{Normalization_Lqq}
N_{qq}\equiv\frac{k_B^2T^3q}{4h}
\end{equation}
is a dimensional factor with $q$ defined in 
(\ref{Scaling_Parameters}).
We find this constraint, which constitutes our first main result, by 
randomly generating transmission matrices obeying the sum rules 
(\ref{Sum_Rules_Transmission_Coefficients})  and exploiting the 
matrix structure (\ref{Landauer_Buettiker_Formula}).

\subsection{Numerical procedure and results}

\begin{figure}
\epsfig{file=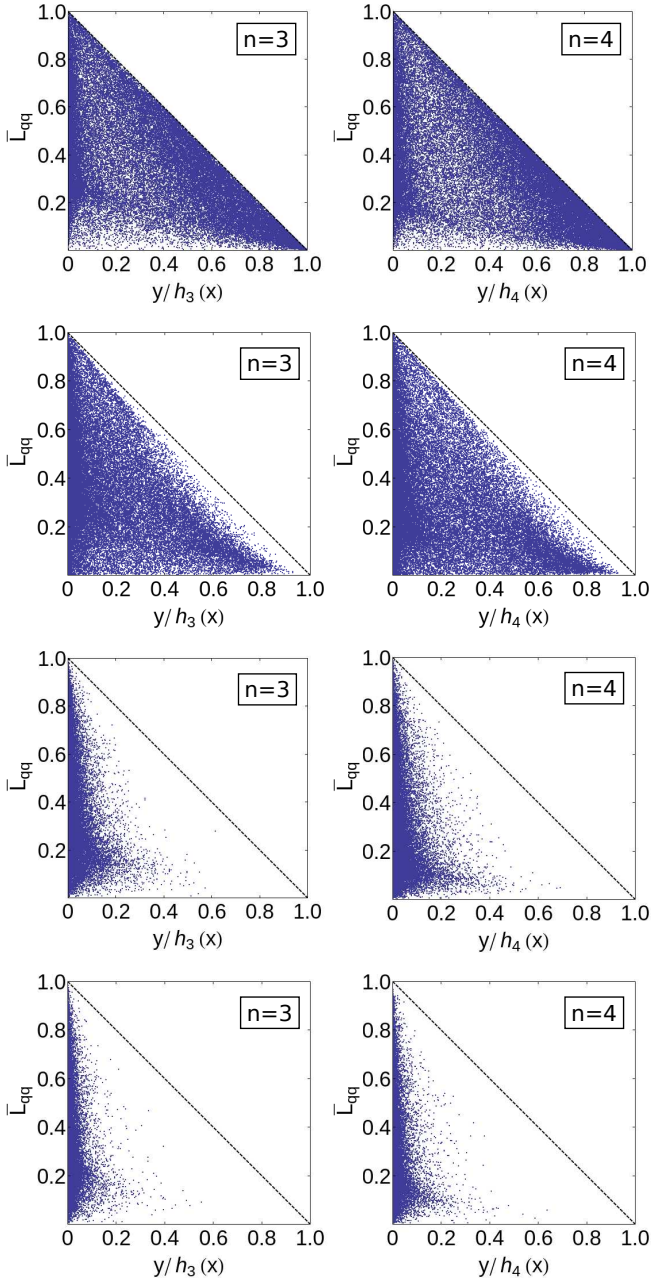,scale=0.9}
\caption{Scatter plots of the rescaled Onsager coefficient 
$\bar{L}_{qq}$ as a function of the ratio $y/h_n(x)$ for 
respectively 50000 randomly chosen $n$-terminal models. 
The dashed line is given by $\bar{L}_{qq}=1-y/h_n(x)$. 
To generate these data, we used the set of permutations 
$\{(1)(2,3), (1,3)(2), (1,3,2)\}$ for $n=3$ (left column) and the 
set $\{(1)(2)(3,4), (1,4)(2,3), (1,4,2,3)\}$ for $n=4$ (right column).
While for all plots the sign of the $w_k$ is chosen randomly, 
$\abs{w_k}$ is sampled from an increasingly large interval 
$\Delta\subseteq [0,1]$. 
Specifically, we have from the first to the fourth line 
$\Delta=[1,1]$, $\Delta=[0.99,1]$, $\Delta=[0.5,1]$, $\Delta=[0,1]$. 
\label{FIG_Point_Plots}}
\end{figure}

\begin{figure}
\epsfig{file=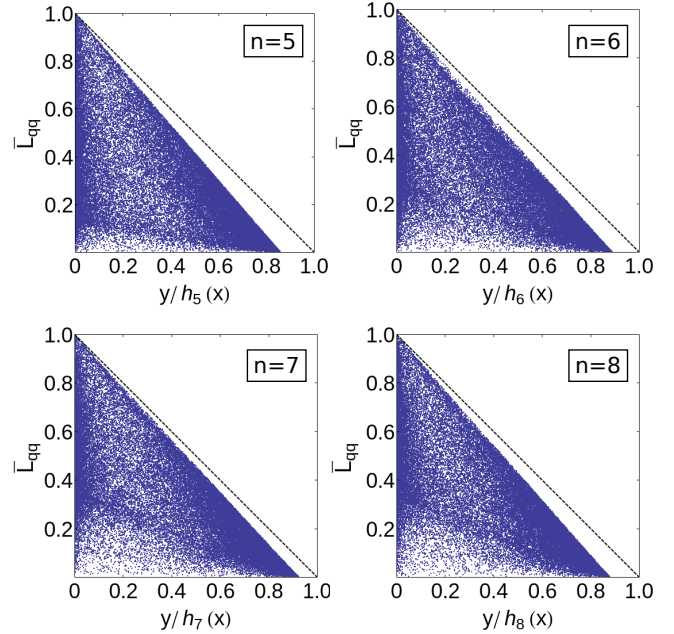,scale=0.9}
\caption{Scatter plots of the rescaled Onsager coefficient 
$\bar{L}_{qq}$ as a function of the ratio $y/h_n(x)$ for 
respectively 50000 randomly chosen $n$-terminal models. 
For all plots, the sign of the $w_k$ is chosen by random and their 
absolute value has been set to $|w_k|=1$.
The sets of permutations used to generate the shown data are
$\{(1,3)(2)(4,5),(1)(2,4)(3,5),(1,4,3,2,5)\}$ for $n=5$, 
$\{(1)(2)(3,6)(4,5),(1,6,4,2,5,3),(1,6,4)(2,3,5)\}$ for $n=6$,
$\{(1,3)(2,4)(5)(6,7),(1,5,7)(2,6)(3,4),(1,7,4,5,2,6,3)\}$ for $n=7$ and
$\{(1,6)(2,8)(3,4,7,5),(1,7,6,3,2,8,5)(4),(1,5,4)$
$(2,3)(6,8)(7)\}$ for $n=8$.
\label{FIG_Point_Plots_2}}
\end{figure}

Our aim for this subsection is to numerically confirm the constraint 
(\ref{Lyh_Bound}). 
Since it is convenient to work with dimensionless 
quantities, we first change the integration variable in 
(\ref{Landauer_Buettiker_Formula}) as $\e\equiv (E-\mu)/(k_B T)$ and, 
second, define the rescaled matrix blocks 
\begin{equation}\label{Rescaled_Matrix_Blocks} 
\begin{split}
\bar{\mathbb{L}}_{ij}  & \equiv
\int_{-\nu}^\infty \!\!\! d\e \; \bar{f}(\e)
\matt{1/p}{\e/\sqrt{pq}}{\e/\sqrt{pq}}{\e^2/q}
\left(\delta_{ij}-\bar{\T}_{ij}(\e)\right)\\
& \equiv \mathbb{D}\mathbb{L}_{ij}\mathbb{D},
\end{split}
\end{equation}
where 
\begin{equation}
\mathbb{D}\equiv\sqrt{\frac{4h}{T}}\left(\!\begin{array}{cc}
1/\sqrt{p} & 0\\ 0 & 1/(k_BT\sqrt{q})
\end{array}\!\right),
\end{equation}
$\bar{f}(\e)\equiv\cosh^{-2}(\e/2)$ and $\bar{\T}_{ij}(\e)\equiv
\T_{ij}(\e k_B T+ \mu)$. 
The scaling parameters $p$ and $q$ are defined by  
\begin{equation}\label{Scaling_Parameters}
\vectt{p}{q} \equiv \int_{-\nu}^\infty d\e \; \bar{f}(\e) \vectt{1}{\e^2}
\xrightarrow{\nu\rightarrow\infty}
\left(\! \begin{array}{c}
4\\
4\pi^2/3
\end{array}\!\right),
\end{equation}
where $\nu\equiv\mu/k_B T$
\footnote{
The explicit evaluation of the integral showing up in 
(\ref{Scaling_Parameters}) for finite $\nu$ yields
\begin{equation*}
\begin{split}
p &= 2\left(1+\tanh(\nu/2)\right)\\
q &= 2\nu\left\{\nu-4\ln(1+e^\nu)+\nu\tanh(\nu/2)\right\}- 8 {{\rm Li}}_2(-e^\nu),
\end{split}
\end{equation*}
where ${{\rm Li}}_2$ denotes the dilogarithm.}.

Next, to obtain a tractable parameterization of the matrix blocks 
(\ref{Rescaled_Matrix_Blocks}), we make use of the fact that 
the matrix $\bar{\mathbb{T}}(\e)$ with elements 
$(\bar{\mathbb{T}}(\e))_{ij}\equiv \bar{\T}_{ij}(\e)$ is bistochastic,
i.e., for any fixed $\e$, its elements fulfill the same sum rules 
(\ref{Sum_Rules_Transmission_Coefficients}) as the $\T_{ij}(E)$.
Consequently, by virtue of the Birkhoff-von Neumann theorem, there 
is a set of $N$ permutation matrices $\mathbb{P}^k$ and positive
numbers $\lambda_k(\e)$ such that 
\begin{equation}\label{Decomposition_Effective_Transmission_Matrix}
\bar{\mathbb{T}}(\e) = \sum_{k=1}^N \lambda_k(\e) \mathbb{P}^k
\quad \text{and} \quad
\sum_{k=1}^N \lambda_k(\e) = 1.
\end{equation}
Inserting this decomposition into 
(\ref{Rescaled_Matrix_Blocks}) and formally carrying out the
integral yields 
\begin{equation}\label{Dimensionless_Matrix_Block_Decomposition}
\bar{\mathbb{L}}_{ij}= \sum_{k=1}^N \left(\delta_{ij}-P^k_{ij}\right)
\cdot \mathbb{M}_k,
\end{equation}
where the $P^k_{ij}\equiv(\mathbb{P}^k)_{ij}$ are the matrix elements 
of the respective permutation matrix and 
\begin{equation}\label{M_Matrix_Integral_Expression}
\mathbb{M}_k\equiv \int_{-\nu}^\infty \!\!\! d\e\;
\bar{f}(\e) \lambda_k(\e)
\matt{1/p}{\e/\sqrt{pq}}{\e/\sqrt{pq}}{\e^2/q}
\end{equation}
is a positive semidefinite, symmetric matrix of dimension $2$.
Due to these properties of the $\mathbb{M}_k$, there exist some 
numbers $\sigma_k, a_k\in (0,\infty)$, $w_k \in [-1,1]$, such that
\begin{equation}\label{M_Matrix_Effective_Parameterization}
\mathbb{M}_k= \sigma_k\matt{1}{w_k a_k}{w_k a_k}{a_k^2}.
\end{equation}
Combining this expression for $\mathbb{M}_k$ with the decomposition
(\ref{Dimensionless_Matrix_Block_Decomposition}) gives a 
parameterization of the dimensionless matrix blocks 
$\bar{\mathbb{L}}_{ij}$ in terms of $N$ permutation matrices 
$\mathbb{P}^k$ of dimension $n$ and $3N$ real parameters $\sigma_k,
a_k, w_k$. 
The latter are constrained by the two important sum rules 
\begin{equation}\label{Sum_Rules_Effective_Parameters}
\sum_{k=1}^N \sigma_k = \sum_{k=1}^N \sigma_k a^2_k = 1,
\end{equation}
which follow directly by comparing 
(\ref{M_Matrix_Effective_Parameterization}) with 
(\ref{M_Matrix_Integral_Expression}) and using the sum rule 
(\ref{Decomposition_Effective_Transmission_Matrix}) for the 
$\lambda_k(\e)$.
We note that, in principle, there is a third sum rule 
\begin{multline}
\sqrt{pq} \sum_{k=1}^N \sigma_k w_k a_k  = \int_{-\nu}^\infty
\!\!\! d\varepsilon\;\bar{f}(\varepsilon) \varepsilon\\
 = \ln (16)
+4\ln\left(\cosh(\nu/2)\right)-2\nu\tanh\left(\nu/2\right).
\end{multline}
However, for simplicity, this constraint will not be exploited
within our analysis. 

We now proceed as follows. 
First, we chose a fixed set of $N$ distinct, $n$-dimensional 
permutation matrices $A_{{\rm P}}\equiv\{\mathbb{P}^k\}_{k=1}^N$ and 
randomly pick a large number of parameter sets $A_{{\rm M}}\equiv
\{\sigma_k,a_k,w_k\}_{k=1}^N$ such that for any of these sets the 
sum rules (\ref{Sum_Rules_Effective_Parameters}) are fulfilled. 
Second, for any of the sets $A_{{\rm M}}$, we evaluate the matrix 
blocks $\bar{\mathbb{L}}_{ij}$ according to
(\ref{M_Matrix_Effective_Parameterization}) and
(\ref{Dimensionless_Matrix_Block_Decomposition}) and subsequently 
calculate the rescaled effective Onsager matrix $\bar{\mathbb{L}}$ 
using (\ref{Effective_Onsager_Matrix}) with the $\mathbb{L}_{ij}$ 
replaced by $\bar{\mathbb{L}}_{ij}$. 
Third, for any individual of the thus obtained matrices 
$\bar{\mathbb{L}}$, we determine the values of the parameters $x$ and
$y$ by inserting the elements of $\bar{\mathbb{L}}$ into the 
definitions given in (\ref{xy_Definition}).
This step is justified, since it is readily seen that the 
effective Onsager matrix $\mathbb{L}$ is connected to its rescaled 
counterpart via the transformation $\mathbb{L}=\mathbb{D}^{-1} 
\bar{\mathbb{L}}\mathbb{D}^{-1}$ and therefore the scaling factors 
contained in $\mathbb{D}$ cancel if the elements of $\mathbb{L}$ 
are plugged into (\ref{xy_Definition}). 
Finally, we plot the rescaled Onsager coefficient $\bar{L}_{qq}$, 
i.e., the lower right entry of the matrix $\bar{\mathbb{L}}$, against
the ratio $y/h_{n}(x)$. 
This choice of variables is quite reasonable, since, as we will show
in the next subsection, the maximum power of the multi-terminal 
thermoelectric generator is proportional to $L_{qq}=N_{qq}
\bar{L}_{qq}$ and we are especially interested in the behavior of 
this quantity as $y$ approaches its bound $h_n(x)$.

We begin with the minimal cases $n=3$ and $n=4$.
Fig. \ref{FIG_Point_Plots} shows the results of the procedure outlined
above for two representative sets $A_{{\rm P}}$ of respectively three 
permutation matrices and all $w_k$ randomly chosen from an 
increasingly large interval. 
We observe that for any of these models the inequality 
(\ref{Lyh_Bound}) is respected and, for some of them, even saturated.
Our numerical data further suggests that this bound is independent of
the choice and number $N$ of distinct permutation matrices 
$\mathbb{P}^k$ in the set $A_{{\rm P}}$, where $N$ must be larger than
two and at least one of the $\mathbb{P}^k$ has to be non-symmetric to 
obtain a non-symmetric effective Onsager matrix. 
However, if $N$ is increased, it becomes more and more improbable to 
find models, for which $\bar{L}_{qq}$ attains the bound 
(\ref{Lyh_Bound}) or comes even close to it. 
Likewise, we find that as the interval $\Delta$, from which the $w_k$
are drawn, is enlarged, the sharp boundary visible in the plots shown
in Fig. \ref{FIG_Point_Plots} deteriorates rapidly.
Therefore, we are convinced that the data presented in the first 
row of Fig. \ref{FIG_Point_Plots} cover a representative subset of
the extreme points in the given parameter space with respect to the 
inequality (\ref{Lyh_Bound}). 

For $n>4$, in Fig. \ref{FIG_Point_Plots_2}, we show representative 
data, which have been obtained for $N=3$ permutation matrices 
$\mathbb{P}_k$ and $|w_k|=1$.
The bound (\ref{Lyh_Bound}) holds for any of the randomly 
chosen models.
However, in contrast to the cases $n=3$ and $n=4$, we were not able 
to achieve saturation. 
Even by increasing the number $N$ beyond $3$, this finding persists.
Sampling the absolute values of the $w_k$ from a finite interval 
$\Delta$, like for $3$ and $4$ terminals, leads to a decay of the
sharp boundary lines appearing in Fig. \ref{FIG_Point_Plots_2}.
It therefore remains an open question at this stage, whether or not,
for $n>4$, there are models that saturate the bound (\ref{Lyh_Bound}).

\subsection{Maximum power}
\begin{figure}
\center
\epsfig{file=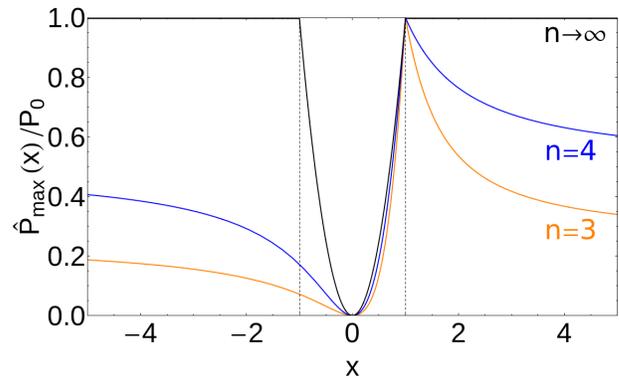,scale=0.25}
\caption{Bound (\ref{Power_Bound}) on the rescaled maximum power 
output $\hat{P}_{{\rm max}}(x)/P_0$ of the multi-terminal 
thermoelectric generator as a function of the asymmetry parameter $x$
for $n=3$, $n=4$ and $n\rightarrow\infty$.
\label{Fig_MaxPower}}
\end{figure}

We will now demonstrate in two steps that the new bound 
(\ref{Lyh_Bound}) significantly restricts the power supplied by a
multi-terminal thermoelectric generator.
First, we maximize (\ref{Power_Output_General}) with respect to
$\gamma$ thus obtaining the maximum power
\begin{equation}\label{Max_Power}
P_{{\rm max}}(x,y) = T (\F_q^{{{\rm h}}})^2 L_{qq} \frac{xy}{4(1+y)}.
\end{equation}
Second, replacing $L_{qq}$ by its upper bound (\ref{Lyh_Bound}), as 
indicated from now on by a hat, and optimizing the resulting 
expression with respect to $y$ gives the upper bound 
\footnote{Remarkably, our new bound (\ref{Power_Bound}) shows the 
same functional dependence on $x$ as the bound 
(\ref{Max_Efficiency_Bound}) on the maximum efficiency.}
\begin{equation}\label{Power_Bound}
\hat{P}_{{\rm max}}(x)\equiv \hat{P}_{{\rm max}}(x,y^\ast(x))
\equiv P_0 x\frac{\sqrt{h_n(x)+1}-1}{\sqrt{h_n(x)+1}+1},
\end{equation}
where
\begin{equation}
P_0\equiv T (\F_q^{{{\rm h}}})^2 N_{qq}/4 
\end{equation}
sets the scale and $y^\ast(x)\equiv\sqrt{h_n(x)+1}-1$.
The bound (\ref{Power_Bound}), which is our second main result,
is plotted in Fig. \ref{Fig_MaxPower}.
In the limit $x\rightarrow\pm\infty$, $\hat{P}_{{\rm max}}$
asymptotically reaches the value $P_0\cos^2(\pi/n)$ and, 
irrespectively of $n$, $\hat{P}_{{\rm max}}(x)$ attains the global 
maximum $P_0$ for $x=1$.  
Moreover, Fig. \ref{Fig_MaxPower} reveals that the bound
(\ref{Power_Bound}) becomes successively weaker as the number of 
terminals is increased. 
In particular for $n\rightarrow\infty$, $\hat{P}_{{\rm max}}(x)$ is 
equal to $P_0$, whenever $|x|\geq 1$.

\subsection{Bounding power by efficiency}
\begin{figure}
\epsfig{file=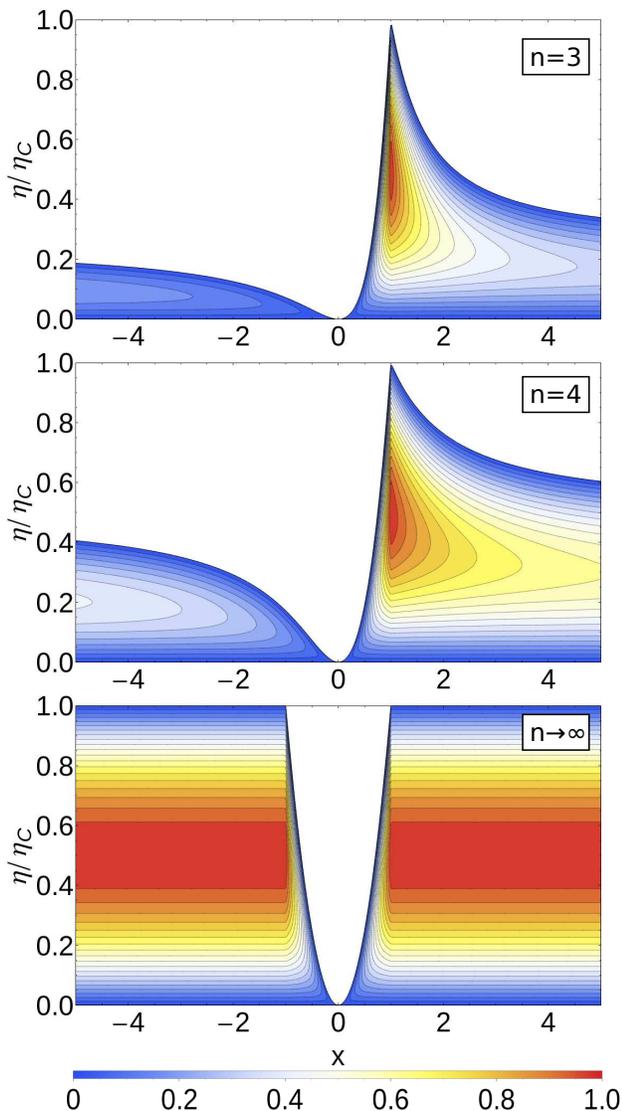,scale=0.49}
\caption{Bounds on the power $\hat{P}_{{\rm max}}(\eta,x)$ of the 
$n$-terminal model as a thermoelectric heat engine in units of $P_0$ 
and as functions of the asymmetry parameter $x$ and the efficiency 
$\eta$ for $n=3,4$ and $n\rightarrow\infty$.
The white regions in the plots are forbidden by the bound 
(\ref{Max_Efficiency_Bound}).
\label{Fig_Power_Efficiency}}
\end{figure}

It is instructive to express the results obtained so far also as a 
function of the efficiency defined in (\ref{Efficiency}).
To this end, we fix the normalized efficiency $\bar{\eta}\equiv\eta/
\eta_{{\rm C}}$ by putting
\begin{equation}\label{Gamma_Fixed_Efficiency}
\gamma = 
\frac{-(\bar{\eta}L_{q\rho}+L_{\rho q})\pm\!\sqrt{(\bar{\eta}L_{q\rho}
+L_{\rho q})^2-4\bar{\eta}L_{\rho\rho}L_{qq}}}{2L_{\rho\rho}}.
\end{equation}
Inserting this $\gamma$ into the general expression 
(\ref{Power_Output_General}) for the power, replacing $L_{qq}$ with 
its upper bound (\ref{Lyh_Bound}) and expressing the result in terms 
of the parameters $x$ and $y$ yields a bound $\hat{P}_\pm(\eta,x,y)$,
whose rather involved expression is given in the appendix. 
By maximizing this function with respect to $y$, we obtain the upper 
bound $\hat{P}_{{{\rm max}}}(x,\eta)$ on power for given efficiency 
$\eta$, asymmetry parameter $x$ and number of terminals $n$, which is 
plotted in Fig. \ref{Fig_Power_Efficiency}.
We find that $\hat{P}_{{{\rm max}}}(\eta,x)$ vanishes 
linearly whenever $\eta$ approaches its previously identified 
upper bound (\ref{Max_Efficiency_Bound}). 
Moreover, in the limit $n\rightarrow\infty$, for which the bound 
$\hat{P}_{{{\rm max}}}(\eta,x)$ becomes the weakest, we end up with
the remarkably simple expression
\begin{equation}\label{Power_Efficiecny_n2Infinity}
\hat{P}_{{{\rm max}}}(\eta,x) = 4 P_0
\begin{cases}
\bar{\eta}(1-\bar{\eta}) & \quad \text{for} \quad \abs{x}\geq 1\\ 
\bar{\eta}-\bar{\eta}^2/x^2 & \quad \text{for} \quad \abs{x}<1
\end{cases},
\end{equation}
which becomes independent of $x$ for $\abs{x}\geq 1$ .
In particular, (\ref{Power_Efficiecny_n2Infinity}) shows 
that the power must vanish at least as $(4P_0/\eta_{{{\rm C}}})
(\eta_{{{\rm C}}}-\eta)$ whenever $\eta$ approaches the Carnot value. 
We can therefore conclude that the new constraint (\ref{Lyh_Bound})
rules out the option of finite power at Carnot efficiency for any 
system that can be described by a multi-terminal model with an 
arbitrary number of probe terminals. 
This insight is our third main result.

\subsection{Power at maximum efficiency}
\begin{figure}
\epsfig{file=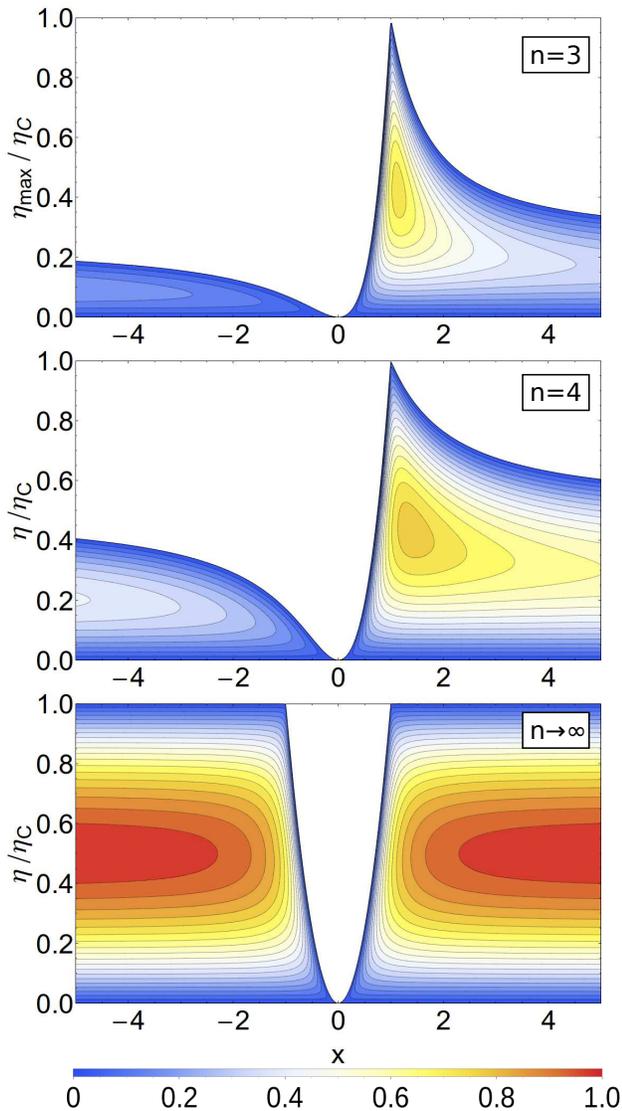,scale=0.49}
\caption{Plots of the bound (\ref{Power@Max_Efficiency}) on the power
at maximum efficiency of a $n$-terminal thermoelectric generator in 
units of $P_0$ and as functions of the asymmetry parameter $x$ and the
maximum efficiency $\eta_{{{\rm max}}}$ for $n=3$, $n=4$ and 
$n\rightarrow\infty$.
\label{Fig_Power@Max_Efficiency}}
\end{figure}

After having determined the maximum power of the $n$-terminal model 
as a heat engine for given efficiency $\eta$, as another quantity of
particular interest, we will now investigate power at maximum 
efficiency. 
To this end, we recall that the maximum efficiency 
(\ref{Max_Efficiency}) of a thermoelectric generator in the linear 
response regime is found by optimizing the general expression 
(\ref{Efficiency}) with respect to $\gamma$ under the condition
$P\geq 0$.
Inserting the thus determined $\gamma$ into 
(\ref{Power_Output_General}) gives the power at maximum efficiency 
\begin{equation}\label{Power@Max_Efficiency}
P^\ast(x,y)=T (\F_q^{{{\rm h}}})^2 L_{qq}  \frac{x}{\sqrt{1+y}}
\frac{\sqrt{1+y}-1}{\sqrt{1+y}+1}.
\end{equation}
By estimating $L_{qq}$ in terms of its upper bound (\ref{Lyh_Bound})
and eliminating $y$ in favor of $\eta_{{\rm max}}$ using 
(\ref{Max_Efficiency}) we obtain the bound
\begin{equation}\label{Bound_PME}
\hat{P}^{\ast}(x,\eta_{{\rm max}})
= 4 P_0\bar{\eta}_{{\rm max}}
\left(\! 
\frac{h_n(x)(x-\bar{\eta}_{{{\rm max}}})^2-4x
\bar{\eta}_{{{\rm max}}}}{h_n(x)(x^2-\bar{\eta}_{{{\rm max}}}^2)}
\!\right)
\end{equation}
on the power output at maximum efficiency, where 
$\bar{\eta}_{{\rm max}}\equiv\eta_{{\rm max}}/\eta_{{\rm C}}$. 

Fig. \ref{Fig_Power@Max_Efficiency} shows plots of the
function $\hat{P}^\ast(x,\eta_{{\rm max}})$ for $n=3$, $n=4$ and 
$n\rightarrow\infty$.
For any finite $n$, this function shows a nontrivial global maximum.
Specifically, we find the maximum $\simeq 0.726 P_0$ at 
$(x,\bar{\eta}_{{\rm max}})\simeq(1.132,0.422)$ for $n=3$ and the 
maximum $\simeq 0.779 P_0$ at $(x,\bar{\eta}_{{\rm max}})\simeq 
(1.373, 0.432)$ for $n=4$. 
For $n\rightarrow\infty$, the global maximum converges to $P_0$.
However, this value can be reached only asymptotically for 
$\bar{\eta}_{{\rm max}}=1/2$ and $x\rightarrow\pm\infty$. 
Moreover, we find that, independent of the number of terminals $n$
and the value of the asymmetry parameter $x$, the power at maximum
efficiency  vanishes like 
$(4 P_0/\eta_{{{\rm C}}})(\eta_{{\rm max}}(x)-\eta_{{\rm max}})$ when
$\eta_{{\rm max}}$ saturates its upper bound 
(\ref{Max_Efficiency_Bound}).

\section{Concluding perspectives}
Strong numerical evidence let us to identify the new constraint 
(\ref{Lyh_Bound}) on the Onsager coefficients describing the coupled 
heat and particle transport in the presence of a magnetic field within
the paradigmatic multi-terminal model for a thermoelectric generator. 
We emphasize that our numerical scheme uses only the sum rules 
(\ref{Sum_Rules_Transmission_Coefficients}) and the structure 
(\ref{Landauer_Buettiker_Formula}) of the matrix blocks the primary
Onsager matrix in (\ref{Phenomenological_Equations}) is composed of.
In particular, the only property of the function $f(E)$ showing up 
in (\ref{Landauer_Buettiker_Formula}) that enters our calculations 
is $f(E)\geq 0$ for any $E\geq 0$. 
Therefore, although here provided for a quantum system, with a 
properly adjusted normalization constant $N_{qq}$ the constraint
(\ref{Lyh_Bound}) and all of its consequences likewise hold for 
classical models like the railway switch model \cite{Horvat2012} and 
the recently proposed classical Nernst engine \cite{Stark} for
which $f(E)$ becomes proportional to the Boltzmann factor.

In the second part, we used the constraint (\ref{Lyh_Bound}) to derive
an upper bound on the power output (\ref{Power_Bound}) as a function 
of the asymmetry parameter $x$ that quantifies the extend to which 
an externally applied magnetic field locally breaks the time reversal 
symmetry of the dynamics inside the scattering region.
In the symmetric case $x=1$ and for $\mu=0$ our bound assumes the 
value $\hat{P}_{{\rm max}}(x=1) = (1/24)\pi^2k_B^2\Delta T^2/h$ with
$\Delta T\equiv T_{{\rm h}}-T$, which is about a factor of $1.3$ 
larger than the bound $P^{{{\rm qb2}}}_{{\rm gen}}\simeq 0.0321 N 
\pi^2 k_B^2 \Delta T^2/h$ for a single conduction channel, i.e., 
$N=1$, which was derived by Whitney \cite{Whitney2014, Whitney2014a}
for a quantum coherent conductor in the nonlinear regime for
vanishing magnetic field. 
The deviation can be explained by the fact that our numerical 
algorithm does not exploit all features of the 
primary Onsager matrix that result from the structure of formula 
(\ref{Landauer_Buettiker_Formula}).
Therefore, the constraint (\ref{Lyh_Bound}) and the bounds it implies 
should not be necessarily regarded as tight, even though they seem to 
be quite good as the comparison with Withney's result for a special 
parameter value shows.

Finally, we analyzed the consequences of (\ref{Lyh_Bound}) 
for the relationship between power and efficiency of the 
multi-terminal model as a thermoelectric generator.
We found that, for any $n$, the maximum power for given $\eta$ must
vanish linearly when $\eta$ approaches its upper bound 
(\ref{Max_Efficiency_Bound}).
Moreover, it turned out that, for $n\rightarrow\infty$, the constraint
(\ref{Lyh_Bound}) implies the bound 
(\ref{Power_Efficiecny_n2Infinity}), which is independent of the 
asymmetry parameter $x$ for $\abs{x}\geq 1$ and comprises the 
bound $\hat{P}_{{{\rm max}}}(\eta,x)$ for any finite $n$. 
Therefore, the assumption that any type of inelastic scattering or 
interaction between electrons can be mimicked by a sufficient 
number of probe terminals attached to the scattering 
region, tempts us to speculate that, up to the normalization constant
$P_0$, the bound (\ref{Power_Efficiecny_n2Infinity}) is universal 
beyond the multi-terminal model.
At the current stage, however, an algebraic proof of (\ref{Lyh_Bound})
within this setup as well as a putative derivation of
(\ref{Power_Efficiecny_n2Infinity}) from first principles without 
reference to a specific model class constitute challenging topics for 
future research.

\appendix*

\begin{widetext}
\section{Calculation of the maximum power for given efficiency}
The function $\hat{P}_\pm(\eta,x,y)$ obtained by the procedure described 
in section III.C after equation (\ref{Gamma_Fixed_Efficiency}) reads 
\begin{equation}
\hat{P}_\pm(\eta,x,y)= 
-4P_0\left(1-\frac{y}{h_n(x)}\right) 
\frac{\bar{\eta}y}{1+y}
\left(
\frac{x+\bar{\eta}}{2x}
-\frac{1+y}{y}\mp\frac{1}{x}\sqrt{\frac{(x+\bar{\eta})^2}{4}
-\frac{x(1+y)\bar{\eta}}{y}}
\right),
\end{equation}
where $\bar{\eta}\equiv \eta/\eta_{{\rm C}}$ denotes the normalized 
efficiency and $y$ is subject to the constraint 
\begin{equation}
\frac{4x\bar{\eta}}{(x-\bar{\eta})^2}\leq y \leq h_n(x)
\quad\text{for}\quad x>0, \qquad
h_n(x)\leq y \leq \frac{4x\bar{\eta}}{(x-\bar{\eta})^2}
\quad\text{for}\quad x<0.
\end{equation}
Since it is readily seen that $P_+(\eta,x,y)\geq P_-(\eta,x,y)$ and we
are interested in an upper bound on power, from here on, we will only 
consider $P_+(\eta,x,y)$. 
With respect to $y$, this function assumes its maximum at 
\begin{equation}\label{yast}
y^\ast(x,\eta) \equiv \bar{\eta}
\frac{2x+(2x-\bar{\eta})h_n(x)+\sqrt{1+h_n(x)}\left(2x +(x-\bar{\eta})h_n(x)\right)}{x^2-2x\bar{\eta}+(\bar{\eta}-x)^2h_n(x)}.
\end{equation}
Consequently, the function $\hat{P}_{{\rm max}}(\eta,x)$ discussed 
in section III.C and plotted in Fig. \ref{Fig_Power_Efficiency}
 is given by 
\begin{equation}\label{Phat}
\hat{P}_{{\rm max}}(\eta,x)\equiv P_+(\eta,x,y^\ast(x,\eta)).
\end{equation}
We note that, in the limit $n\rightarrow\infty$, (\ref{yast}) reduces
to 
\begin{equation}
\frac{4x\bar{\eta}}{(x-1)(1+x-2\bar{\eta})}
\quad\text{for}\quad \abs{x}\geq 1, \qquad
\frac{4\bar{\eta}}{x-x^3-2\bar{\eta}+2x\bar{\eta}}
\quad\text{for}\quad \abs{x}<1
\end{equation}
and (\ref{Phat}) assumes the simple form 
(\ref{Power_Efficiecny_n2Infinity}).
\end{widetext}



%

\end{document}